\documentclass[epj]{svjour}
\usepackage{epsfig}
\usepackage{url}

\newcommand{\X}{\ensuremath{\langle X \rangle}}
\newcommand{\pvec}{\ensuremath{\mathbf{p}}}
\newcommand{\Estar}{\ensuremath{E^*_{\gamma}}}
\newcommand{\qstar}{\ensuremath{\theta^*_{\gamma}}}
\newcommand{\dTCA}{\ensuremath{d_{\mathrm{TC}}}}
\newcommand{\dtTCA}{\ensuremath{d_{\mathrm{TC}\bot}}}
\newcommand{\dc}{\ensuremath{d_\mathrm{PCA}}}
\newcommand{\lc}{\ensuremath{l_\mathrm{PCA}}}
\newcommand{\rxy}{\ensuremath{r_{xy}}}
\newcommand{\cv}{\ensuremath{\mathbf{x}_{\rm C}}}
\newcommand{\nv}{\ensuremath{\mathbf{x}_{\rm N}}}
\newcommand{\dnc}{\ensuremath{d_{\rm NC}}}
\newcommand{\Emiss}{\ensuremath{E_{\rm miss}}}
\newcommand{\pvecmiss}{\ensuremath{\mathbf{p}_{\rm miss}}}
\newcommand{\pmiss}{\ensuremath{p_{\rm miss}}}

\newcommand{\kl}{\ensuremath{K_L}}
\newcommand{\ks}{\ensuremath{K_S}}
\newcommand{\keg}{\ensuremath{K^0_{e3\gamma}}}
\newcommand{\kegf}{\ensuremath{K^0_{e3}}}
\newcommand{\klele}{\ensuremath{K_L\to\pi^+e^-\overline{\nu}}}
\newcommand{\klpos}{\ensuremath{K_L\to\pi^-e^+\nu}}
\newcommand{\klpen}{\ensuremath{K_L\to\pi e\nu}}
\newcommand{\klpmn}{\ensuremath{K_L\to\pi\mu\nu}}
\newcommand{\kspp}{\ensuremath{K_S\to\pi^+\pi^-}}
\newcommand{\klpp}{\ensuremath{K_L\to\pi^+\pi^-}}
\newcommand{\klppp}{\ensuremath{K_L\to\pi^+\pi^-\pi^0}}

\newcommand{\ChPT}{\mbox{$\chi$PT}}
\newcommand{\DAFNE}{DA\char8NE}
\newcommand{\GeV}{\mbox{GeV}}
\newcommand{\ie}{i.e.}
\newcommand{\MeV}{\mbox{MeV}}
\newcommand{\stat}{\ensuremath{_{\rm stat}}}
\newcommand{\syst}{\ensuremath{_{\rm syst}}}

\newcommand{\ab}{\ensuremath{\sim}}
\newcommand{\order}[1]{\ensuremath{\mathcal{O}(#1)}}
\newcommand{\SN}[2]{\ensuremath{#1\times10^{#2}}}

\newcommand{\Ref}[1]{Ref.~\cite{#1}}
\newcommand{\Refs}[1]{Refs.~\cite{#1}}
\newcommand{\andRef}[1]{and~\cite{#1}}
\newcommand{\Eq}[1]{Eq.~(\ref{#1})}
\newcommand{\Eqs}[1]{Eqs.~(\ref{#1})}
\newcommand{\andEq}[1]{and~(\ref{#1})}
\newcommand{\Fig}[1]{Fig.~\ref{#1}}
\newcommand{\Tab}[1]{Table~\ref{#1}}
\newcommand{\Sec}[1]{Sec.~\ref{#1}}

\begin{document}

\title{
A study of the radiative \boldmath{$K_L \to \pi^\pm e^\mp \nu \gamma$}
decay and search for direct photon emission with the KLOE detector}
\titlerunning{
A study of the radiative $K_L \to \pi^\pm e^\mp \nu \gamma$ decay}

\subtitle{The KLOE Collaboration}

\author{
F.~Ambrosino\inst{5,6} \and
A.~Antonelli\inst{1} \and
M.~Antonelli\inst{1} \and
F.~Archilli\inst{10,11} \and
C.~Bacci\inst{12,13} \and
P.~Beltrame\inst{2} \and
G.~Bencivenni\inst{1} \and
S.~Bertolucci\inst{1} \and
C.~Bini\inst{8,9} \and
C.~Bloise\inst{1} \and
S.~Bocchetta\inst{12,13} \and
F.~Bossi\inst{1} \and
P.~Branchini\inst{13} \and
R.~Caloi\inst{8,9} \and
P.~Campana\inst{1} \and
G.~Capon\inst{1} \and
T.~Capussela\inst{1} \and
F.~Ceradini\inst{12,13} \and
F.~Cesario\inst{12,13} \and
S.~Chi\inst{1} \and
G.~Chiefari\inst{5,6} \and
P.~Ciambrone\inst{1} \and
F.~Crucianelli\inst{8} \and
E.~De Lucia\inst{1} \and
A.~De Santis\inst{8,9} \and
P.~De Simone\inst{1} \and
G.~De Zorzi\inst{8,9} \and
A.~Denig\inst{2} \and
A.~Di Domenico\inst{8,9} \and
C.~Di Donato\inst{6} \and
B.~Di Micco\inst{12,13} \and
A.~Doria\inst{6} \and
M.~Dreucci\inst{1} \and
G.~Felici\inst{1} \and
A.~Ferrari\inst{1} \and
M.L.~Ferrer\inst{1} \and
S.~Fiore\inst{8,9} \and
C.~Forti\inst{1} \and
P.~Franzini\inst{8,9} \and
C.~Gatti\inst{1} \and
P.~Gauzzi\inst{8,9} \and
S.~Giovannella\inst{1} \and
E.~Gorini\inst{3,4} \and
E.~Graziani\inst{13} \and
W.~Kluge\inst{2} \and
V.~Kulikov\inst{16} \and
F.~Lacava\inst{8,9} \and
G.~Lanfranchi\inst{1} \and
J.~Lee-Franzini\inst{1,14} \and
D.~Leone\inst{2} \and
M.~Martini\inst{1,7} \and
P.~Massarotti\inst{5,6} \and
W.~Mei\inst{1} \and
S.~Meola\inst{5,6} \and
S.~Miscetti\inst{1} \and
M.~Moulson\inst{1} \and
S.~M\"uller\inst{1} \and
F.~Murtas\inst{1} \and
M.~Napolitano\inst{5,6} \and
F.~Nguyen\inst{12,13} \and
M.~Palutan\inst{1} \and
E.~Pasqualucci\inst{9} \and
A.~Passeri\inst{13} \and
V.~Patera\inst{1,7} \and
F.~Perfetto\inst{5,6} \and
M.~Primavera\inst{4} \and
P.~Santangelo\inst{1} \and
G.~Saracino\inst{5,6} \and
B.~Sciascia\inst{1} \and
A.~Sciubba\inst{1,7} \and
A.~Sibidanov\inst{1} \and
T.~Spadaro\inst{1} \and
M.~Testa\inst{8,9} \and
L.~Tortora\inst{13} \and
P.~Valente\inst{9} \and
G.~Venanzoni\inst{1} \and
R.~Versaci\inst{1,7} \and
G.~Xu\inst{1,15}
}

\institute{
Laboratori Nazionali di Frascati dell'INFN, Frascati, Italy \and
Institut f\"ur Experimentelle Kernphysik, Universit\"at Karlsruhe, Germany \and
Dipartimento di Fisica dell'Universit\`a, Lecce, Italy \and
INFN Sezione di Lecce, Lecce, Italy \and
Dipartimento di Scienze Fisiche dell'Universit\`a ``Federico II'', 
Napoli, Italy \and
INFN Sezione di Napoli, Napoli, Italy \and
Dipartimento di Energetica dell'Universit\`a ``La Sapienza'', Roma, Italy \and
Dipartimento di Fisica dell'Universit\`a ``La Sapienza'', Roma, Italy \and
INFN Sezione di Roma, Roma, Italy \and
Dipartimento di Fisica dell'Universit\`a ``Tor Vergata'', Roma, Italy \and
INFN Sezione di Roma Tor Vergata, Roma, Italy \and
Dipartimento di Fisica dell'Universit\`a ``Roma Tre'', Roma, Italy \and
INFN Sezione di Roma Tre, Roma, Italy \and
Physics Department, State University of New York, Stony Brook, USA \and
Institute of High Energy Physics of Academica Sinica, Beijing, China \and
Institute for Theoretical and Experimental Physics, Moscow, Russia
}
 
\authorrunning{The KLOE Collaboration}

\mail{Marco.Dreucci@lnf.infn.it}

\date{Received: date / Revised version: date}

\abstract{
We present a measurement of the ratio
$R = \Gamma(\keg;\Estar>30~\MeV,\qstar>20^\circ)$/$\Gamma(\kegf)$ 
and a first measurement of the direct emission contribution in semileptonic
\kl\ decays. The measurement was performed at the \DAFNE\ $\phi$ factory 
by selecting $\phi\to\kl\ks$ decays with the KLOE detector.
We use 328 pb$^{-1}$ of data, corresponding to about 3.5 million \kegf\ events 
and about 9000 \keg\ radiative events. Our result is 
$R = \SN{(924 \pm 23\stat \pm 16\syst)}{-5}$ for the branching ratio 
and $\X = -2.3 \pm 1.3\stat \pm 1.4\syst$ for the effective strength 
parameter describing direct emission.}


\maketitle

\section{Introduction}
\label{sec:intro}
The study of radiative \kl\ decays provides information about the structure
of the kaon and the opportunity to quantitatively test theories 
describing hadron interactions and decays, such as chiral perturbation 
theory (\ChPT). 
In addition, the correct understanding of radiation
in \kl\ decays is necessary for precision measurements of the 
fully-inclusive decay rates. These, in turn, are needed for
studies of the decay dynamics and the determination 
of the CKM matrix element $|V_{us}|$.  

Two different processes 
contribute to photon emission in kaon decays: inner bremsstrahlung (IB) 
and direct emission (DE). 
DE is radiation from intermediate hadronic states and is sensitive 
to hadron structure.
The relevant kinematic variables for the study of radiation in $K_{\ell3}$
decays are \Estar, the energy of the radiated photon, and \qstar,
its angle with respect to the lepton momentum in the kaon rest frame.
The IB amplitudes diverge for $\Estar \to 0$. For $K_{e3}$, for which
$m_e \approx 0$, the IB spectrum in \qstar\ is peaked near zero as well.
The IB and DE amplitudes interfere.
The contribution to the width from IB-DE interference is 1\% or less of the 
purely IB contribution; the purely DE contribution is negligible.
To disentangle
the two components, we measure the double differential rate
$d^2\Gamma/d\Estar\,d\qstar$.

In the \ChPT\ treatment of Ref.~\cite{kubis}, DE is 
characterized by eight amplitudes, \{$V_i$, $A_i$\}, 
which in the one-loop approximation are real 
functions. These terms have similar photon energy
spectra, with maxima around $\Estar = 100~\MeV$. 
This suggests a decomposition of the photon spectrum at \order{p^6} as
in \Ref{kubis}:
\begin{eqnarray}
  \frac{d\Gamma}{d\Estar} & = & \frac{d\Gamma_{\rm IB}}{d\Estar} + 
  \sum_{i=1}^4 \left( \langle V_i \rangle \frac{d \Gamma_{V_i}}{d\Estar} + 
  \langle A_i \rangle \frac{d \Gamma_{A_i}}{d\Estar} \right) \nonumber \\
                          & \simeq &
  \frac{d\Gamma_{\rm IB}}{d\Estar} + \X f(\Estar).
  \label{eq:x}
\end{eqnarray}
The DE contributions are summarized in the function $f(\Estar)$, 
which represents the deviation from the IB spectrum. 
The parameter \X\ measures the effective strength of the DE.
\order{p^6} \ChPT\ calculations give 
$\X = -1.2 \pm 0.4$, a $3\sigma$ indication that the IB-DE 
interference is destructive.\footnote
{The quantity \X\ was evaluated for $\qstar>5^\circ$ in \Ref{kubis},
instead of $\qstar>20^\circ$ as in the present analysis. It turns
out \cite{private} that this makes a negligible difference.}
The low-energy constants (LECs) for the \order{p^6} terms
are unknown. An educated guess of their size leads to the assignment
of an uncertainty on \X\ of 30\% of the \order{p^4} 
result \cite{kubis}.

A first attempt to measure the DE contribution was performed by the KTeV 
collaboration \cite{KTeV:DE} using the model described in \Refs{ffs,doncel} 
within the so-called soft-kaon approximation.
However, as shown in \Ref{kubis}, in this approximation there is insufficient
sensitivity for the evaluation of the contribution from DE.
In contrast, our fit to the double differential 
spectrum $d^2\Gamma/d\Estar\,d\qstar$ allows us to isolate DE from IB.

We also measure the ratio $R$, conventionally defined as
\begin{equation}
  R \equiv \frac{\Gamma(\keg;\Estar>30~\MeV,\qstar>20^\circ)}{\Gamma(\kegf)},
  \label{eq:ratio}
\end{equation}
where $\Gamma(\kegf)$ represents the decay width inclusive of 
radiative effects. The value of this ratio has been computed at 
\order{p^6} in \ChPT, leading to the prediction \cite{private} 
\begin{equation}
R = \SN{(0.963 + 0.006\,\X \pm 0.010)}{-2}.
\label{eq:rx}
\end{equation}
For $\X = -1.2$, $R = \SN{(0.96\pm0.01)}{-2}$, as quoted in \Ref{kubis}.
The simultaneous measurement of $R$ and $\X$ allows a precise 
comparison with the theory, in large part avoiding complications
from the uncertainties on the LECs for \order{p^6}.

In 2001, KTeV published \cite{KTeV:DE}
the result $R = \SN{(0.908\pm0.008^{+0.013}_{-0.012})}{-2}$;
the data were subsequently
reanalyzed using more restrictive cuts that provide better control over
systematic effects, but which reduce the statistics by a factor of three.
The more recent KTeV result \cite{KTeV:R} is
$R = \SN{(0.916\pm0.017)}{-2}$. 
In 2005, NA48~\cite{NA48:R} measured 
$R = \SN{(0.964\pm0.008^{+0.011}_{-0.009})}{-2}$.
Neither of these experiments measure \X.

\section{Experimental setup}
The data were collected with the KLOE detector at \DAFNE, the Frascati
$\phi$ factory. \DAFNE\ is an $e^+e^-$ collider that operates at 
a center of mass energy of \ab1020~\MeV, the mass of the $\phi$ meson.
Positron and electron beams of equal energy collide at an angle of 
($\pi-0.025$) rad, producing $\phi$ mesons with a small momentum 
in the horizontal plane ($p_\phi \ab13$~\MeV).
$\phi$ mesons decay \ab34\% of the time into nearly collinear \ks\kl\ pairs; 
the detection of a \ks\ (the tagging kaon) therefore signals the 
presence of a \kl\ (the tagged kaon), independently of the decay mode
of the latter.
This principle is called \kl\ tagging in the following.

The KLOE detector consists of a large cylindrical drift chamber
surrounded by a lead/scintillating-fiber electromagnetic 
calorimeter.
A superconducting coil around the calorimeter provides a 0.52~T field. 
The drift chamber \cite{DC} is 4~m in diameter and 3.3~m long.
The momentum resolution for tracks at large polar angles is 
$\sigma_{p_{\perp}}/p_{\perp}\approx 0.4\%$.
The vertex between two intersecting tracks is reconstructed with a 
spatial resolution of \ab3~mm.
The calorimeter \cite{EMC} is divided into a barrel and two endcaps. 
It is segmented in depth into five layers and covers 98\% of the solid angle.
Energy deposits nearby in time and space are grouped into calorimeter 
clusters. The energy and time resolutions are 
$\sigma_E/E = 5.7\%/\sqrt{E\ (\GeV)}$ and
$\sigma_T = 57~{\rm ps}/\sqrt{E\ (\GeV)}\oplus100~{\rm ps}$, respectively.
For this analysis, the trigger~\cite{TRG} uses only calorimeter information.
Two energy deposits above threshold ($E>50$~\MeV\ for the barrel and  
$E>150$~\MeV\ for endcaps) are required.
Recognition and rejection of cosmic-ray events is also performed at the 
trigger level. Events with two energy deposits above a 30~\MeV\ threshold 
in the outermost calorimeter plane are rejected.

The 328 pb$^{-1}$ of data used in this analysis were collected 
in 2001 and 2002. The data are divided into 14 run periods of about 
25~pb$^{-1}$/period. For each data period, 
we have a corresponding sample of Monte Carlo (MC) 
events with approximately equivalent statistics.

\section{Monte Carlo generators}
The KLOE MC generates only radiation from IB, so
a dedicated generator for DE is needed.
Moreover, the accuracy of the KLOE IB generator is a relevant issue.
The KLOE generator \cite{gatti} uses
a resummation in the soft-photon limit to all orders in $\alpha$
of the \order{p^2} amplitude for single photon emission.
It describes the IB photon spectrum at the level of 
\ab1\%, which is appropriate for inclusive decay-rate measurements
at the 0.1\% level.
However, since the DE contribution is about 1\% of the IB contribution,
the accuracy level of the KLOE IB generator is of about the same order as 
the DE contribution itself.
From the point of view of the measurement of $R$, this could introduce an 
error of only \ab1\%. 
On the other hand, a fit-based counting procedure making use of an IB
distribution biased by \ab1\% could introduce a \ab100\% error in
the number of \keg\ events from DE. 
Therefore, in this analysis, we use the generator of \Ref{kubis} 
to describe the photon spectrum from IB as well as from DE.
This generator is based on an \order{p^6} calculation; 
the code was provided by the authors.
The generator is incorporated into the KLOE MC and reconstruction 
program.
This is the first analysis of the double differential spectrum to make
use of an \order{p^6} generator.

\section{Analysis}
\label{sec:analysis}
The criteria used to select an inclusive sample of \kegf\ events 
are the same described in \Ref{kloe:ff}. We briefly summarize them here.

Candidate \kl\ events are tagged by the presence of a \kspp\ decay.
The tagging efficiency is independent of \Estar\ and \qstar. Over the
range of \Estar\ the efficiency fluctuates around 66\% with an rms of 0.3\%;
over the range of \qstar\ the rms fluctuation is 0.1\%.

We search for a \kl\ decay along the direction of the \kl\ momentum as 
reconstructed from the \kspp\ decay ({\it{tagging line}}). 
All tracks in the chamber, after removal of those from the \ks\ 
decay and their descendants, are extrapolated to their points of closest
approach (PCA) to the tagging line. 
For each track candidate, we evaluate the distance \dc\ of closest approach
to the tagging line.
The length of extrapolation of the track to this point of closest approach,
\lc, is also computed.
Tracks satisfying $\dc < a\rxy + b$, with $a=0.03$ and $b=3$ cm, and 
$-20 < \lc <25$~cm are accepted as \kl\ decay products, where \rxy\ 
is the distance of the vertex from the origin in the transverse plane. 
For each sign of charge we consider the track with the smallest value 
of \dc\ to be associated to the \kl\ decay. 
Starting from these track candidates, a two-track vertex is reconstructed.
An event is retained if the vertex is in the fiducial volume 
$35 < \rxy < 150$~cm and $|z| < 120$~cm. The tracking and vertex efficiencies 
are evaluated by MC simulation and corrected using data control 
samples \cite{kloe:ff,KLOE:BR}.

To remove background from \klppp\ and \klpp\ decays with 
minimal efficiency loss, we apply loose kinematic cuts. 
Assuming the two tracks to have the pion mass,  
we require $\Emiss^2 - \pmiss^2 - M_{\pi^0}^2 < -5000~\MeV^2$ and 
$\sqrt{\Emiss^2 + \pmiss^2} > 10~\MeV$, where \Emiss\ and \pvecmiss\
are the missing energy and momentum, respectively.
A large amount of background from \klpmn\ decays is rejected using
the variable $\Delta_{\pi\mu}$, the lesser value of $|\Emiss - \pmiss|$
calculated in the two hypotheses, $\pi\mu$ or $\mu\pi$.
We retain events only if this variable is greater than 10~\MeV.

These kinematic criteria do not provide enough suppression of the
background from \klpen\ decays with incorrect track-particle assignment 
and from \klpmn\ decays.
We make use of time-of-flight (TOF) information from the calorimeter
to further reduce the contamination \cite{kloe:ff}.   
For each \kl\ decay track with an associated cluster, we define the variable: 
$\Delta t_i = t_{\rm clu}-t_{i} ,~(i = \pi,~ e) $ in which $t_{\rm clu}$ is 
the cluster time and $t_{i}$ is the expected time of flight, evaluated 
according to a well-defined mass hypothesis.
An effective way to select the correct mass assignment, $\pi e$ or $e\pi$,
is obtained by choosing the lesser of $|\Delta t_{\pi^+}-\Delta t_{e^-}|$ and
$|\Delta t_{\pi^-}-\Delta t_{e^+}|$.
After the mass assignment has been made, we consider the variables
$\Delta t_{\pi}+\Delta t_e$ and $\Delta t_{\pi}-\Delta t_e$. 
We select the signal by using a $2\sigma$ cut, where the resolution 
$\sigma\simeq 0.5~\mbox{ns}$.
We take the TOF efficiency from the Monte Carlo after correcting the time 
response of the calorimeter using data control samples~\cite{KLOE:kssemi}.

For the purposes of track-to-cluster association, 
we define two quantities related to the distance between the 
extrapolation of the track to the calorimeter entry point
and the nearest cluster: 
\dTCA, the distance from the extrapolated entry point to 
the cluster centroid, and \dtTCA, the component of
this distance in the plane transverse to the momentum of the track
at the entry position. We only consider clusters with $\dtTCA < 30$~cm.
We evaluate the clustering efficiency using the Monte Carlo, and correct it
with the ratio of data and Monte Carlo efficiencies obtained from control
samples \cite{kloe:ff}.

The inclusive \kegf\ reconstruction efficiency is about 0.25 and differs
by \ab6\% for \klele\ and \klpos\ events (see \Ref{kloe:ff}).
We therefore count the number of \kegf\ events separately for each charge.
In all, we find about 3.5 million \kegf\ events with a
contamination of \SN{7}{-3} mainly due to \klppp\ and \klpmn\
decay events.

We select signal \keg\ events from within the inclusive \kegf\ sample.
We first search for events with a photon cluster, \ie, a calorimeter
cluster not associated with any track.
Assuming that the \kl\ decay vertex lies on the tagging line,
the arrival time of each photon gives an independent determination of the
\kl\ decay position, \nv, the so-called neutral vertex.
The method is fully described in \Refs{EMC} \andRef{offline}. 
We require that the distance \dnc\ between the position 
\nv\ of the neutral vertex and the
position \cv\ of the \kl\ vertex determined by track reconstruction, 
to be within eight times the rms of the \dnc\ distribution for MC signal 
events.
If there is more than one photon candidate, we choose the one with
the smallest value of \dnc.
We retain events reconstructed with $\qstar>20^\circ$.

To evaluate the photon energy we use the track momenta and the 
photon cluster position. Specifically, we write for the photon momentum
\begin{equation}
  \pvec_\gamma = 
     E_\gamma\:\frac{\mathbf{x}_{\rm clu}-\nv}
                    {|\mathbf{x}_{\rm clu}-\nv|}, 
  \label{eq:eneg}
\end{equation}
and for the missing four-momentum
\begin{equation}
  p_\nu = p_K - p_\pi - p_e - p_\gamma,
  \label{eq:four}
\end{equation}
where $p_\nu$, $p_K$, $p_\pi$, $p_e$, and $p_\gamma$ are the particle
four-momenta. 
Setting $p_\nu^2=0$ and solving the above equations
gives $E_\gamma$, the photon energy in the laboratory system,
with a resolution of \ab1~\MeV. This resolution is about a factor of ten 
better than that obtained using the energy measurement from the calorimeter.

\begin{figure}
  \centering
  \psfig{figure=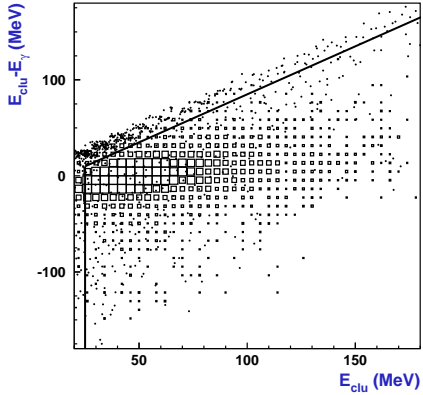,width=6.0cm}
  \caption{Cut used to remove events with accidental activity (dots) 
           from the sample of signal \keg\ events (boxes),
           in MC simulation.}
  \label{fig:effgsela}
\end{figure}
\begin{figure}
  \centering
  \psfig{figure=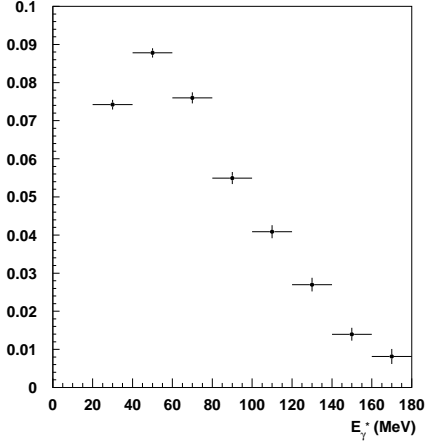,width=6.0cm}
  \caption{\keg\ signal efficiency, relative to the sample of tagged $K_L$ 
           decays, from MC.}
  \label{fig:effgselb}
\end{figure}
The main background contribution at this stage is from \keg\ events with 
an undetected soft photon to which a cluster from machine background 
has been accidentally associated. This background is strongly reduced by 
requiring $E_{\rm clu}>25$~\MeV, where $E_{\rm clu}$ is the cluster energy 
as measured in the calorimeter, and by cutting on the difference between
$E_{\rm clu}$ and $E_\gamma$ as shown in \Fig{fig:effgsela}.
We reduce the relative contribution from background by a factor of
four, with \ab7\% loss in signal efficiency.

The background contributions from \klppp\ and \klpmn\ events 
after application of the above cuts are 4.2\% and 2.5\%, 
respectively. The reconstructed photon energy and angular distributions 
for background events from these sources overlap with those for DE events. 
We use neural network techniques to reduce these backgrounds. 
To remove \klppp\ events, we use a neural network based on the 
photon energy and angle (with respect to the momentum of the lepton candidate), 
the track momenta, the missing momentum, and $M^2_{\gamma\nu}$, 
the invariant mass of the photon-neutrino pair. 
To remove \klpmn\ events, we use a neural network based on the 
track momenta, the calorimeter energy measurement, and the 
cluster centroid position.
Cuts on the neural network output reduce background from 
\klppp\ decays from 4.2\% to 0.4\%, and from 
\klpmn\ decays, from 2.5\% to 1.4\%. 
The signal loss is 10\%.
Figure~\ref{fig:effgselb} shows the selection efficiency
for signal events after all cuts, relative to the inclusive \kegf\ sample,
as evaluated by MC. Averaged over the spectrum of \Estar, 
the absolute efficiency for detection of a \keg\ event from IB 
is $(6.92\pm0.04)\%$. Because the \Estar\ spectrum for DE events is 
harder, the average absolute detection efficiency in this case is 
slightly lower: $(5.80\pm0.05)\%$.

To check the data-MC agreement, calibrate the MC position \nv, and 
correct the photon-selection efficiency, we use \klppp\ decays 
as a control sample. 
These events are selected using a tight cut in the variable 
$\Emiss^2 - \pmiss^2 - m_{\pi^0}^2$, evaluated assigning 
the pion mass to both tracks.
We additionally require the presence of a cluster with $E_{\rm clu}>60$~\MeV\ 
not associated to any track, 
corresponding to one of the two photons from $\pi^0$ decay. 
This high-energy photon is used to 
tag the presence of the second photon. 
We select about $350\,000$ \klppp\ events with a purity of 
99.8\%.

We first compare the resolution for photon energy reconstruction
in data and MC.
We reconstruct the energy of the second photon using \Eqs{eq:eneg}
\andEq{eq:four}, where the tagging photon is ignored and plays the 
role of the undetected neutrino in signal events.
\begin{figure}
   \centering
      \psfig{figure=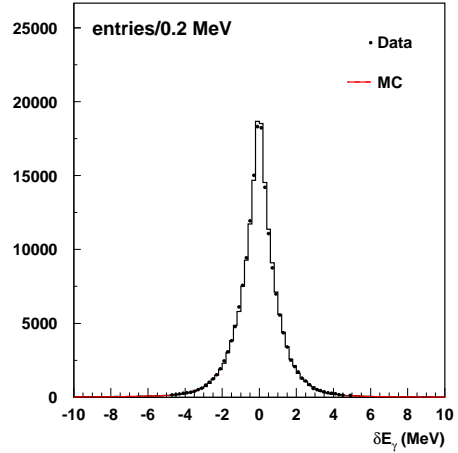,width=6cm}
   \caption{Comparison of resolution for photon energy reconstruction
            for \klppp\ events in data and MC.} 
   \label{fig:gammacsa}
\end{figure}
\begin{figure}
   \centering
      \psfig{figure=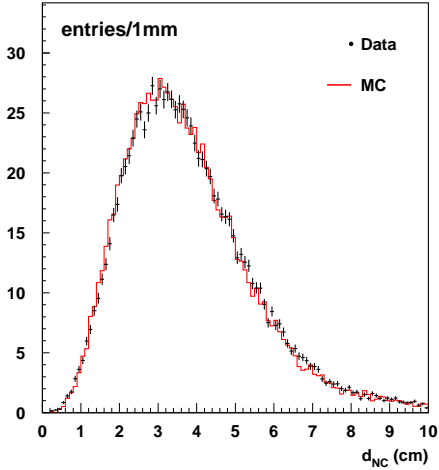,width=6cm}
   \caption{Comparison of \dnc\ distributions for data and
            MC for \klppp\ vertices in the central part of the drift chamber,
            after corrections.}
   \label{fig:gammacsb}
\end{figure}
The expected value of the energy of the second photon is computed 
using the reconstructed momentum of the tagging photon to close the 
kinematics at the decay vertex.
Figure~\ref{fig:gammacsa} shows the distribution of the residuals
for both data and MC. Good agreement is observed.

In addition, we use the \klppp\ control sample to check the reconstruction 
of the distance \dnc\ and its rms in both data and MC 
in order to tune the MC simulation (see \Fig{fig:gammacsb}).

We also use the \klppp\ sample to validate the MC simulation of the 
calorimeter energy response, since we apply analysis cuts on $E_{\rm clu}$ 
to remove accidentals. The energy scale is about 2~\MeV\ lower in MC
than in data. To good approximation, this bias is
independent of energy.

Finally, we use the \klppp\ sample to evaluate the photon selection
efficiency for data and MC. We obtain a correction of a few percent,
which we apply to the simulation.
Further details can be found in \Ref{KLOE:ke3gnote}.

\section{Fit} 
\label{sec:fit}
\begin{figure*}
   \centering
   \psfig{figure=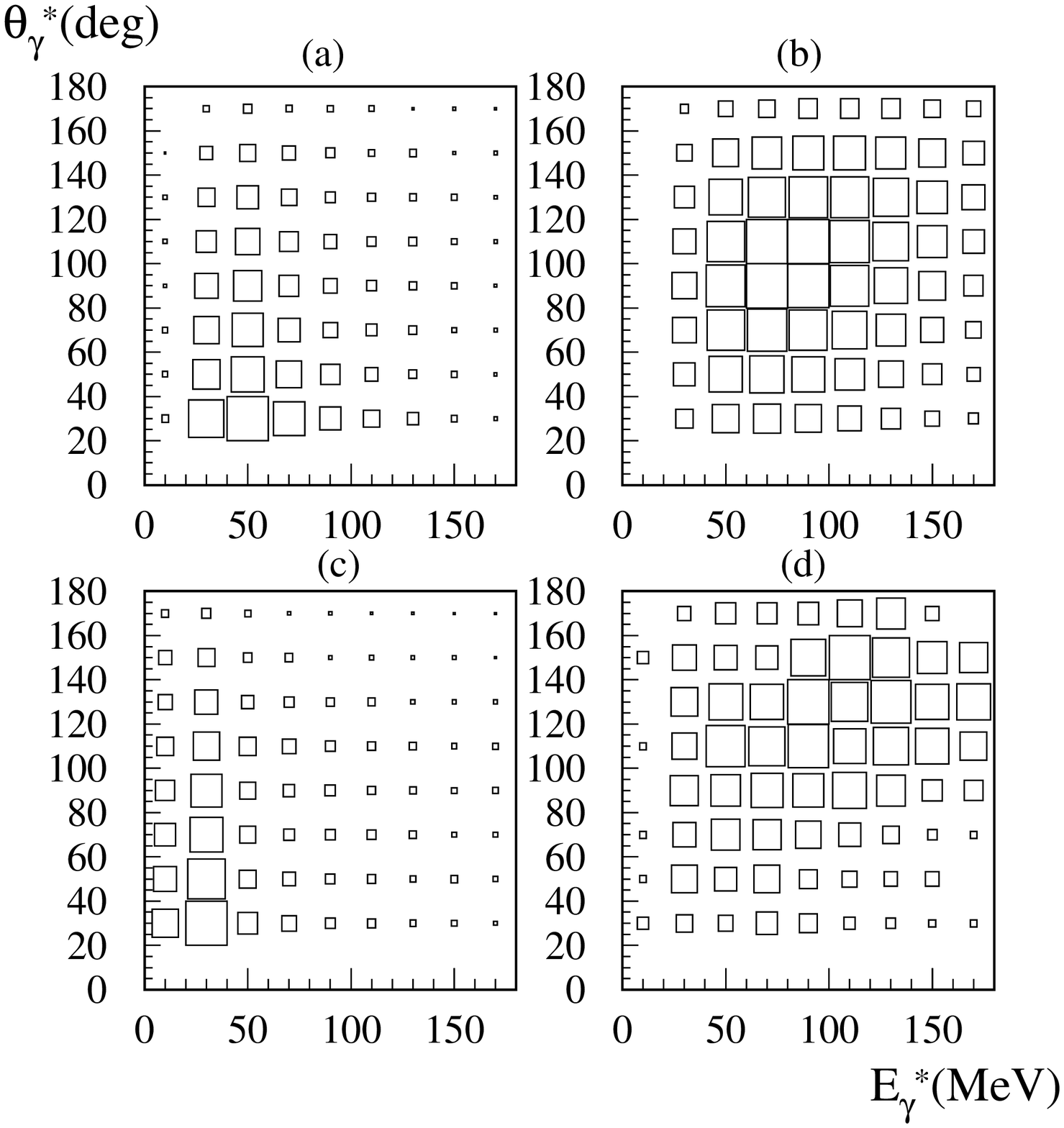,width=10cm}
   \caption{Reconstructed Monte Carlo distributions in \qstar\ (deg) 
     vs.\ \Estar\ (MeV). 
     From top left: (a) \keg\ events from IB,
                    (b) \keg\ events from DE as defined in the text,
                    (c) \keg\ events from IB not satisfying 
                        the \Estar\ and/or \qstar\ cuts as generated,
                    (d) background events from \klppp\ and \klpmn.
     The statistics for different plots do not respect natural proportions.}
   \label{fig:4plot}
\end{figure*}
\begin{figure*}[t!]
   \centering
   \psfig{figure=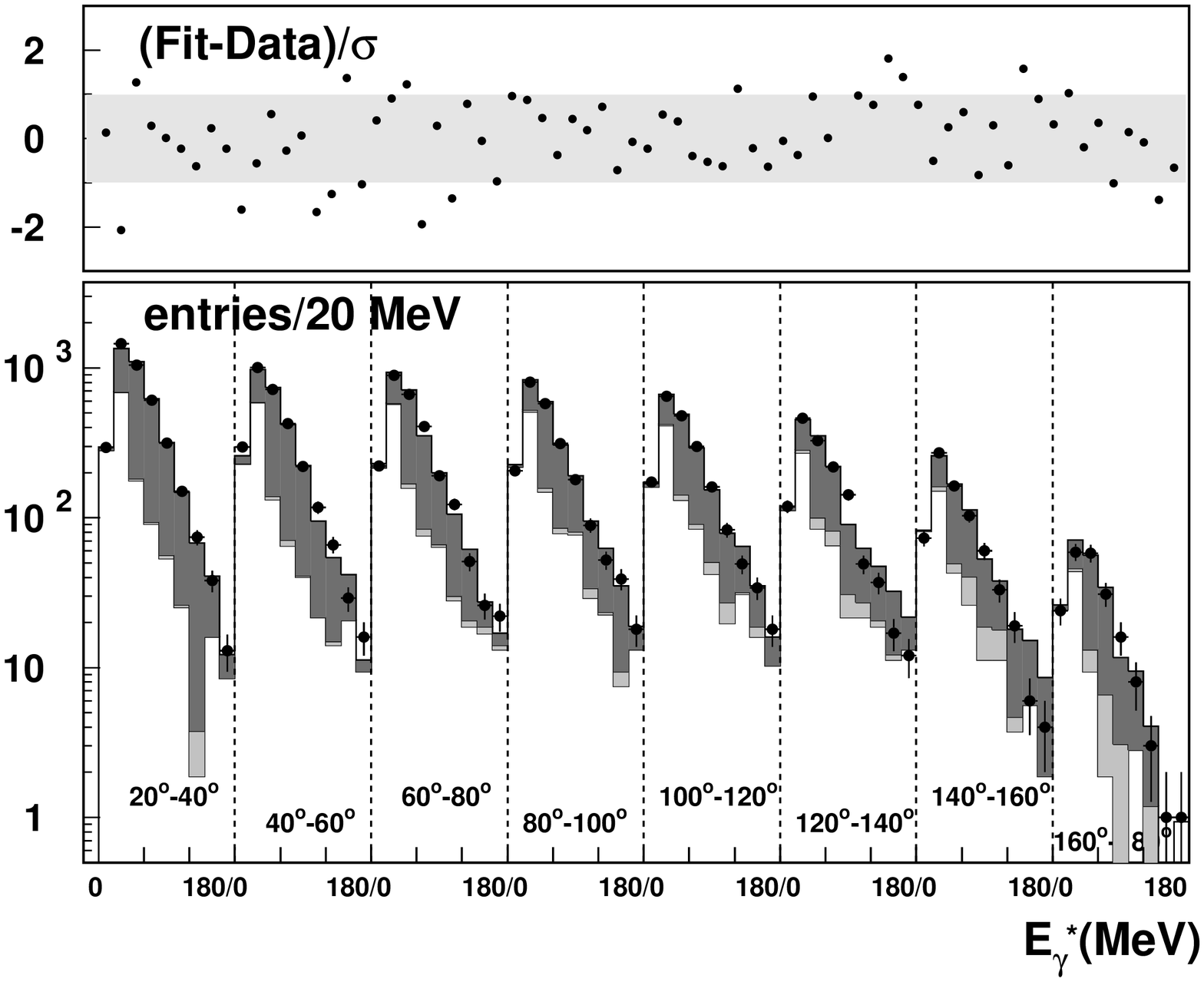,width=12cm}
   \caption{Results of fit to $(\Estar,\qstar)$ distribution: 
          dots show data, dark gray region shows contribution from 
          \keg\ events (IB and DE) satisfying kinematic cuts, 
          white region shows contribution from 
          \keg\ events not satisfying cuts,
          light gray region shows contribution from \klppp\ and \klpmn\ decays.
          Above: Normalized fit residuals.}
   \label{fig:fit}
\end{figure*}

We perform a fit to the experimental distribution in $(\Estar,\qstar)$ 
using the sum of four independently normalized MC distributions:
\begin{itemize} 
\item the distribution for \keg\ events from IB satisfying the kinematic 
      cuts $\Estar>30$~\MeV\ and $\qstar>20^\circ$ as generated; 
\item the distribution corresponding to the function $f(\Estar)$ in the
      second term of \Eq{eq:x}, representing the modification of the 
      spectrum from DE events satisfying the kinematic cuts as generated; 
\item \keg\ events from IB \emph{not} satisfying the kinematic cuts as 
      generated;
\item physical background from \klppp\ and \klpmn\ events.
\end{itemize} 
These four MC distributions are shown in \Fig{fig:4plot}. 
The free parameters of the fit are the number of IB events,
the effective number of DE events (the integral of the spectral 
distortion induced by the IB-DE interference), 
and the number of \keg\ events not satisfying the kinematic cuts.
We fix the background contribution from \klppp\ and \klpmn\ using the MC.
Figure~\ref{fig:fit} shows the result of the fit. 
The two-dimensional distributions are plotted on a single axis; 
the \Estar\ distributions for each of the eight slices in \qstar\ 
are arrayed sequentially.
The eight slices in \qstar\ are $20^\circ$ each and cover the interval
from $20^\circ$ to $180^\circ$.
The values obtained for the fit parameters are listed, 
together with their correlation coefficients, 
in \Tab{tab:count}. The fit gives 
$\chi^2/{\rm ndf}=60/69$ ($P = 77\%$).
\begin{table}[hbt]
  \begin{center}
    \caption{Values obtained for fit parameters, with correlations.}
    \begin{tabular}{lcc|ccc}
      \hline 
      Contribution      &  $N$   & $\delta N$  & 
                                  \multicolumn{3}{c}{Correlation coeffs.} \\
      \hline
      IB                & 9083   & 213 & 1 & & \\
      \keg\ not in cuts & 6726   & 194 & $-0.586$ &\\
      DE (effective)    & $-102$ & 59  & $-0.254$ & $-0.022$ & 1\\
      \hline
    \end{tabular}
    \label{tab:count}
  \end{center}
\end{table}
The negative value for the effective number of counts from DE events 
is a result of the destructive interference between the IB and DE amplitudes.
The presence of DE modifies the total number of \keg\ events satisfying 
the kinematic cuts at the level of \ab1\%. 
From the fit results, we obtain
\begin{eqnarray*}
  R & \equiv & \frac{\Gamma(\keg;\Estar>30~\MeV,\qstar>20^\circ)}
                    {\Gamma(\kegf)}\\
    & = & \SN{(924 \pm 23\stat)}{-5}.
\end{eqnarray*}

\section{Systematic uncertainties} 
We estimate systematic uncertainties by varying the selection cuts.
Signal events are defined by the tracking, clustering, track-to-cluster 
association, neutral-vertex acceptance, and analysis cuts.
Any variation of these cuts produces a variation in the result.

{\bf Tagging}\hspace{1em}
In obtaining our result, we do not require that the tagging \kspp\ decay 
by itself satisfy the calorimeter trigger.
This requirement may be imposed by demanding the identification of two
clusters that are associated to tracks from the \kspp\ decay and which 
fire trigger sectors.
Doing this makes the analysis independent of the MC estimate of the trigger
efficiency, at a cost in statistics. When we impose this requirement as a 
check, we observe a variation $\Delta R = \SN{4}{-5}$.

{\bf Tracking}\hspace{1em}
The most selective variable in the definition of track candidates is \dc, 
the distance of closest approach of the track to the 
tagging line. As described in \Sec{sec:analysis}, we accept tracks 
with $\dc < a\rxy + b$, with $a=0.03$ and $b=3$ cm.
\dc\ is reconstructed with a resolution of about 1~cm.
The tracking efficiency depends most sensitively on the value of $b$.
We vary $b$ from 2 to 5~cm and re-evaluate the run-period-dependent
tracking efficiency correction in each case. 
The uncertainty on the tracking efficiency correction is dominated
by sample statistics. We observe a variation in the result
$\Delta R = \SN{1.5}{-5}$. The width of the \lc\ distribution is $\sim$4~cm, 
so that the cut on \lc\ is quite loose ($\sim$$5\sigma$), and we assign
no corresponding contribution to the systematic error.   

{\bf Clustering}\hspace{1em}
The most selective variable used for track-to-cluster association
is the transverse distance \dtTCA. This distance is reconstructed with 
a resolution of about 6~cm. We vary the cut on \dtTCA\ from 
15 to 50~cm, around a nominal value of 30~cm. 
For each value of \dtTCA, we re-evaluate the clustering efficiency correction, 
which is run-period dependent. Here also, the uncertainty in the correction 
is dominated by sample statistics.
We observe a variation $\Delta R = \SN{5.5}{-5}$.

{\bf Kinematic cuts}\hspace{1em}
We apply loose kinematic cuts. When these cuts are varied, 
the variation in the result is negligible.

{\bf TOF cuts}\hspace{1em}
TOF cuts are used in the identification of the inclusive \kegf\ sample.
When the TOF cut is varied by 30\% around its nominal $2\sigma$ value,
we observe a variation in the result $\Delta R = \SN{1.3}{-5}$.

{\bf Momentum miscalibration and resolution}\hspace{1em}
We have also considered effects from the momentum scale accuracy and 
resolution.
We assume a maximum momentum scale uncertainty of 0.1\% \cite{offline},
which corresponds to a variation in the result $\Delta R = \SN{3.5}{-5}$.
Changing the value assumed for the momentum resolution by $\pm3\%$
as in \Ref{kloe:ff}
gives rise to a variation $\Delta R = \SN{7.2}{-5}$.

{\bf Fiducial volume}\hspace{1em}
Reducing the fiducial volume by 20\% produces a variation in the result
$\Delta R = \SN{3}{-5}$.

{\bf Rejection of accidentals.}\hspace{1em}
The cut used to remove accidentally associated background 
clusters is illustrated in \Fig{fig:effgsela}.
The cut is tightest at lower energies.
At $E_{\rm clu} = 25$~MeV, we require
$E_{\rm clu}-E_\gamma < 10$~MeV, 
while for signal events the variance
of this residual is \ab8~MeV.
Varying the intercept of this cut
by $\pm5$~\MeV\ gives rise to a variation
in the result $\Delta R = \SN{5.2}{-5}$.

{\bf Neutral-vertex acceptance}\hspace{1em}
We search for a neutral vertex within a sphere centered around \cv.
We accept events for which \dnc\ is less than eight times the rms 
of the \dnc\ distribution for signal events.  
We vary the cut from six to ten times the rms of the distribution
and observe a variation in the result $\Delta R = \SN{2.9}{-5}$.

{\bf Background}\hspace{1em}
Conservatively, we remove the cuts on the neural network outputs. 
This increases the background level by nearly a factor of four. 
The variation in the result is 
$\Delta R = 9\times10^{-5}$.

{\bf Fit systematics}\hspace{1em}
As a check, we perform the fit leaving free the number of background 
events from \klppp\ and \klpmn.
This gives consistent results, but with a greater statistical uncertainty. 
In particular, the total number of background events from the fit is
$406\pm152$, as compared to the MC expectation, $301\pm17$.
We have also checked the fit stability as a function of run period. 
This requires fixing the number of background events, because within a 
single run period, the background 
statistics are too low to guarantee good fit convergence. In addition, 
we do not include DE, as the data from a single run period offer 
no sensitivity to this component. 
The stability over run periods is good: a fit to determine the average 
value of $R$ gives $\chi^2/{\rm ndf} = 9/13$. Therefore, we assign no
contribution to the systematic uncertainty from this source.

\begin{table}[hbt]
  \begin{center}
    \caption{Summary of the absolute systematic uncertainties on $R$ and 
      $\X$.}
    \begin{tabular}{lcc}
      \hline
      Source              &  $10^5\times\Delta R$ & $\Delta\X$ \\
      \hline
      Tagging             &    4.0   &  0.7     \\
      Tracking            &    1.5   &  0.8     \\  
      Clustering          &    5.5   &  0.1     \\
      TOF cut             &    1.3   &  0.5     \\
      $p$ miscalibration  &    3.5   &  0.2     \\
      $p$ resolution      &    7.2   &  0.4     \\
      Fiducial volume     &    3.0   &  0.5     \\
      Accidentals         &    5.2   &  0.4     \\
      Neutral vertex acceptance       &    2.9   &  0.3     \\
      Background          &    9.0   &  0.1     \\
      \hline
      Total               &    15.5  &  1.4     \\
      \hline
    \end{tabular}  
    \label{tab:syst}    
  \end{center}
\end{table}
All systematic errors are summarized in \Tab{tab:syst}.
These errors are added in quadrature to obtain the final systematic error.

\section{Results} 
Our final result for $R$ is
\begin{displaymath}
  R = \SN{(924 \pm 23\stat \pm 16\syst)}{-5}.
\end{displaymath}
\begin{figure}
   \centering
      \psfig{figure=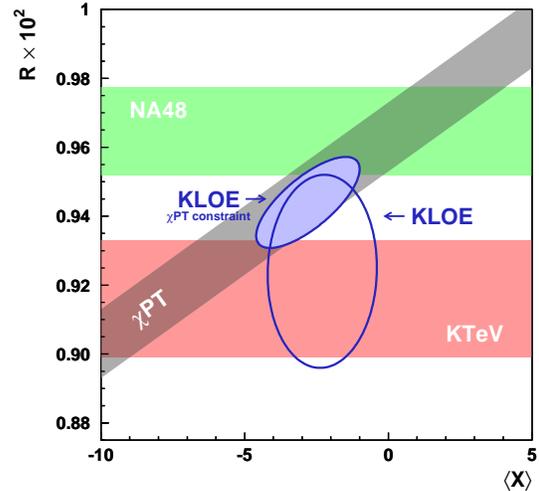,width=7cm}
   \caption{KLOE $1\sigma$ contours in the  $(R,\X)$ plane from fit 
            to the $(\Estar,\qstar)$ distribution (open ellipse), and 
            same results when combined with constraint from
            \ChPT\ (filled ellipse). Results from KTeV and NA48 are also
            shown, as well as the dependence of $R$ on \X\ according to
            \Eq{eq:rx}, used as the constraint.} 
   \label{fig:rx}
\end{figure}
The value of the parameter \X\ defined in \Eq{eq:x} is derived from 
the result of the fit that gives the effective numbers of IB and DE 
events, taking into account the difference in the overall detection 
efficiencies for each type (the detection efficiency for IB events is 
\ab20\% higher than for DE events).
We obtain
\begin{displaymath}
  \X = -2.3 \pm 1.3\stat \pm 1.4\syst. 
\end{displaymath}
The systematics on \X\ are evaluated in the same manner as 
for $R$, and the different contributions are listed in 
\Tab{tab:syst}.
The correlation coefficient between the total errors on $R$ and \X\ 
is 3.9\%. The $1\sigma$ contour is illustrated in \Fig{fig:rx}.

The dependence of $R$ on \X\ of \Eq{eq:rx} is shown in \Fig{fig:rx} as 
the diagonal shaded band. This dependence can be used to further constrain
the possible values of $R$ and \X\ from our measurement, giving 
the $1\sigma$ contour illustrated as the filled ellipse 
in the figure. The constraint is applied via a fit, which gives
$R = \SN{(944\pm14)}{-5}$ and
$\X = -2.8\pm1.8$, 
with correlation $\rho = 72\%$ and 
$\chi^2/{\rm ndf} = 0.64/1$ ($P = 42\%$).
This result represents an improved test of \ChPT\ with respect to that 
obtained using the measurements of $R$ from \Refs{KTeV:R} \andRef{NA48:R}.

Finally, to test the accuracy of the \order{p^2} KLOE IB 
generator \cite{gatti}, we have performed fits to the data 
with no DE component. We obtain
\begin{equation}
\begin{array}{ll}
\mbox{\order{p^6} generator \cite{kubis}}\quad 
    & R = \SN{(925\pm23\stat)}{-5}; \\
\mbox{\order{p^2} generator \cite{gatti}}\quad
    & R = \SN{(921\pm23\stat)}{-5}. \\
\end{array}
\end{equation}
The fit with the \order{p^6} generator gives $\chi^2/{\rm ndf} = 63/70$ 
($P=71\%$); that with the \order{p^2} generator gives
$\chi^2/{\rm ndf} = 68/70$ ($P=55\%$).
The agreement between these results confirms the reliability of the KLOE
generator for IB events.

\section{Conclusion} 
Two different components contribute to photon emission in \keg\ decays: IB
and DE. The latter describes photon radiation from intermediate hadronic
states, providing additional information on the hadronic structure of the
kaon. \ChPT\ predicts that the IB and DE amplitudes interfere, resulting in
a negative effective strength \X. 

From a fit to the $(\Estar,\qstar)$ distribution for \keg\ decays
based on \order{p^6} \ChPT\ calculations, we obtain 
a value for $R$ and a first measurement of \X.
These results, which
favor destructive interference between the IB and DE amplitudes,
are good agreement with the \ChPT\ predictions.

\begin{acknowledgement}
We would like to thank Bastian Kubis, one of the authors of \Ref{kubis}, 
for providing the Monte Carlo generator used in this analysis.
We thank the \DAFNE\ team for their efforts in maintaining 
low-background running conditions and their collaboration during 
all data taking. 
We want to thank our technical staff: 
G.F.~Fortugno and F.~Sborzacchi for their dedicated work to ensure
efficient operation of the KLOE Computing Center; 
M.~Anelli for his continuous support to the gas system and the safety of
the detector; 
A.~Balla, M.~Gatta, G.~Corradi and G.~Papalino for maintenance of the
electronics;
M.~Santoni, G.~Paoluzzi and R.~Rosellini for general support to the
detector; 
C.~Piscitelli for his help during major maintenance periods.
This work was supported in part
by EURODAPHNE, contract FMRX-CT98-0169; 
by the German Federal Ministry of Education and Research (BMBF),
contract 06-KA-957; 
by the German Research Foundation (DFG), 'Emmy Noether Programme'
contracts DE839/1-4;
by INTAS, contracts 96-624 and 99-37; 
and by the EU Integrated Infrastructure
Initiative HadronPhysics Project, contract
RII3-CT-2004-506078.
\end{acknowledgement}

\end{document}